\documentclass[aps,twocolumn,showpacs,fleqn,nofootinbib]{revtex4}
\usepackage[utf8]{inputenc}
\usepackage{amsmath}
\usepackage{graphicx}
\usepackage{caption}
\usepackage{subcaption}
\usepackage{float}
\usepackage{verbatim}
\usepackage{color}
\usepackage{comment}
\usepackage{physics}
\usepackage{booktabs}

\usepackage{hyperref}

\def\lsim{\mathrel{\rlap{\lower4pt\hbox{\hskip1pt$\sim$}}
    \raise1pt\hbox{$<$}}}                % less than or approx. symbol
\def\gsim{\mathrel{\rlap{\lower4pt\hbox{\hskip1pt$\sim$}}
    \raise1pt\hbox{$>$}}}                % greater than or approx. symbol

\begin{document}
	
\title{Investigating QCD Dynamical Entropy in high-energy nuclear collisions}
\author{G.S. Ramos $^{1}$} 
\email{silveira.ramos@ufrgs.br}
\author{L.S. Moriggi$^{2}$}
\email{lucasmoriggi@unicentro.br}
\author{M.V.T. Machado$^{1}$}
\email{magnus@if.ufrgs.br}

\affiliation{$^{1}$ High Energy Physics Phenomenology Group, GFPAE. Institute of Physics, Federal University of Rio Grande do Sul (UFRGS)\\
Caixa Postal 15051, CEP 91501-970, Porto Alegre, RS, Brazil} 
\affiliation{$^{2}$ Universidade Estadual do Centro-Oeste (UNICENTRO), Campus Cedeteg, Guarapuava 85015-430, Brazil}

\begin{abstract}
In this work, the concept of QCD dynamical entropy is extended to heavy ion systems. This notion of entropy can be understood as a relative entropy and can also be used to estimate the initial entropy density in ultra-relativistic heavy ion collisions. The key quantity used to calculate this entropy is the nuclear unintegrated gluon distribution (nUGD), which provides a transverse momentum probability density. In the numerical analysis, both the geometric scaling phenomenon and the Glauber-Gribov approach have been used to evaluate realistic models for the nUGD. It is shown that the normalization procedure and the geometric scaling property make the dynamical entropy almost independent of the nucleus mass number $A$. Results are presented for the dynamical entropy density, $dS_D/dy$, in terms of the rapidity.
\end{abstract}	

\maketitle
	
\section{Introduction}
\label{sec1}

In recent years, statistical mechanics tools have been applied in high-energy physics \cite{Munier:2009pc,Iancu:2004es,Le:2022exz}. For example, if one considers that a hadron wavefunction can be divided into two entangled subsets, where the first one is measured by deep inelastic scattering (DIS) and its counterpart remains unmeasured, an entanglement entropy can be computed \cite{Kharzeev:2017qzs,Kharzeev:2021nzh}. Using the color dipole framework, the entanglement entropy is expressed in terms of the gluon density, and phenomenological works show good agreement with proton-proton ($pp$) \cite{Hentschinski:2024gaa,Hentschinski:2023izh,Hentschinski:2022rsa,Zhang:2021hra,Kharzeev:2021yyf,H1:2020zpd,Ramos:2020kaj} and electron-proton ($ep$) \cite{Tu:2019ouv,Gotsman:2020bjc,Germano:2021brq,Ramos:2024,Kutak:2022} collision data. Another example is the entropy produced in proton-proton and heavy-ion collisions, which is related to the particle multiplicities generated in these processes.

In this way, the statistical physics approach is relevant when assuming a transient evolution of the quark-gluon plasma (QGP). Nonetheless, a theoretical difficulty exists in relating its properties to the basic concepts of QCD. In this conundrum, studies using strong gauge coupling employ the Anti-de Sitter/Conformal Field Theory (AdS/CFT) correspondence \cite{Maldacena:2000}, starting from multiple initial conditions. This approach leads the thermalization phase to the hydrodynamic one in a strongly-coupled gauge theory plasma, depending on an initial entropy factor. Therefore, the entropy released in these collisions may be related to the difference between the final hydrodynamic entropy and the initial one, when the QGP is in a QCD weak-coupling initial state in the context of the Color Glass Condensate (CGC), also known as the glasma (pre-equilibrium) stage. The concept of dynamical entropy aims to define this initial entropy \cite{Peschanski:2013}. In particular, it gives a microscopic definition of entropy analogous to the Boltzmann approach, based on the QCD dynamics of the CGC.

As defined in Ref. \cite{Peschanski:2013}, the QCD dynamical entropy is a sort of  relative entropy (RE) or Kullback–Leibler divergence \cite{10.2996/kmj/1138844604,Floerchinger:2020ogh}.  From the classical point of view, the relative entropy can be understood as a
measure of distinguishability between two distributions $P$ and $Q$, with $S(P\parallel Q)=\sum_j P_j\ln (P_j/Q_j)$.  $S(P\parallel Q)$ is a non-negative quantity that is zero if and only if the two distributions are equal. The relative entropy has some advantages over the usual entropy, such as: i) it is well-defined for
discrete and continuous random variables and ii) it is invariant under a reparameterization of coordinates  on the underlying statistical manifold. In quantum field theory, the quantum relative entropy (QRE) between two states represented by the reduced density matrices $\rho$ and $\sigma$ is defined as $S(\rho\parallel \sigma)=\mathrm{Tr}[\rho (\ln \rho - \ln \sigma)]$    \cite{Vedral:2002zz,Nielsen:2012yss}. In this case, the QRE is named relative entanglement entropy \cite{Berges:2017hne}. Thus, QRE acts as a measure of distinguishability of two states. In the context of quantum information theory, the distinguishability of different physical states quantified by relative entropy is a key ingredient for information processing.

In a previous work \cite{Ramos2022}, we focused on the phenomenology of the dynamical entropy $\Sigma^{Y_0\rightarrow Y}$ of dense gluonic states in proton-proton collisions. The total rapidity is written in terms of Bjorken $x$, $Y=\ln (1/x)$, with the notation $Y_0 = \ln (1/x_c)$ for the initial total rapidity. Here, $x_c\sim 10^{-2}$ is the expected limit of validity of the QCD color dipole approach.  The dynamical entropy aims to quantify the amount of disorder caused by the non-linear rapidity evolution $Y_0\rightarrow Y$ of the CGC medium. From a physical perspective, this entropy can be evaluated in two complementary ways: macroscopically and microscopically. The first approach is related to its derivation from far-from-equilibrium thermodynamics. In particular, the Jarzynski-Crooks identity \cite{Jarzynski:1997,Crooks:1999} relates the distribution of thermodynamic work in the process $A\rightarrow B$ to the Helmholtz free energy $\Delta F$ between the equilibrium states in the process $A\rightarrow C$, while the Hatano-Sasa identity \cite{Hatano:2001} generalizes the former to non-equilibrium states \cite{Mounier:2012}.

The microscopical point of view of QCD dynamical entropy is related to the balance between branching and recombination in the saturation regime of dense QCD states. In this regime, gluons organize themselves into cells of typical saturation scale size $R_s(Y) = 1/Q_s(Y)$, where $Q_s \sim e^{\lambda Y}$ is the momentum saturation scale at a total rapidity $Y$, with $\lambda$ fitted. From this, the non-linear energy evolution can be evaluated as a compression $R_s(Y_0) \rightarrow R_s(Y) < R_s(Y_0)$ for $Y_0 < Y$.

The QCD dynamical entropy is evaluated through the transverse momentum probability distributions $P(k,Y)$, where $k$ is the gluon transverse momentum. These distributions are obtained via a normalization procedure of the unintegrated gluon distributions $\varphi(k,Y)$ (UGD). In the work of Ref. \cite{Ramos2022}, we calculated these distributions using three phenomenological UGDs: the Gaussian Golec-Biernat and Wusthoff (GBW) model \cite{GolecBiernat:1998,Kutak:2011}; the Tsallis statistics inspired  Moriggi, Peccini, and Machado (MPM) model \cite{Moriggi:2020zbv,Moriggi:2020qla}; and the UGD based on the Levin-Tuchin (LT) model for the saturated region \cite{Abir:2017mks,Siddiqah:2018qey}.

From this point of view, this work is a direct continuation of the dynamical entropy studies, but now, focused on proton-nucleus ($pA$) collisions. The reason is that we can compare the microscopic entropy interpretation given by the dynamical entropy with the  notion of a thermodynamical entropy associated with the gluon production in the saturation regime of CGC initial states in nuclear collisions in the dilute-dense regime \cite{Kutak:2011rb}.  The key feature is to obtain a nuclear UGD, although these distributions are scarce in the literature. To achieve this, we will adapt the proton UGD using two different strategies. The first strategy centers on one of the main properties of high energy saturation physics: geometric scaling. In a deep inelastic scattering, during the $\gamma^*A$ interaction, geometric scaling ensures that the atomic mass number, $A$ dependence of the ratio of the DIS cross-section to the hadron area $\pi R^2_h$ can be absorbed into the $A$-dependence of the momentum saturation scale $Q_{s,A}^2(Y)$ \cite{Armesto:2004}. The second strategy is based on the Glauber-Gribov approach \cite{Glauber:1959}, where, ignoring isospin effects that are negligible in the small-$x$ region, the total dipole-proton cross-section $\sigma_{dp}(x,r)$, with $r$ as the dipole size, is replaced by its nuclear form, $\sigma_{dA}(x,r)=\int d^2b \, \sigma_{dA}(x,r,b)$, where $b$ is the impact parameter and $\sigma_{dA}(x,r,b)$ is the total dipole-nucleus cross-section for a fixed $b$. 

The plan of the paper is as follows: In the next section [\ref{sec2}], we will present the basic concepts of QCD dynamical entropy, including the transverse momentum distribution functions $P(k,Y)$, the dynamical entropy $\Sigma^{Y_0\rightarrow Y}$, and its density $dS_D/dy$. We will also discuss both strategies mentioned above, geometric scaling and the Glauber-Gribov approach. In section \ref{sec3}, we will present the main results on the dynamical entropy observables. Finally, in section \ref{sec4}, a summary of our findings is provided.

\section{Theoretical formalism}
\label{sec2}

In our first step, it is necessary to briefly review the formulation of the dynamical entropy of dense QCD states proposed in Ref. \cite{Peschanski:2013}. These states are described in weakly coupled QCD that evolves nonlinearly from the dilute partonic state. The evolution with total rapidity $Y$ increases the parton density to the saturation limit, where it can be described as a CGC state. In the initial state, one has an initial total rapidity $Y_0$, associated with a transverse size $R_s(Y_0)=1/Q_s(Y_0)$, with $Q_s$ being the rapidity-dependent saturation scale that evolves to $Y$, with $R_s(Y)$. The saturation scale is given by $Q_s^2(x)=Q_0^2(x_0/x)^\lambda$ or $Q_s^2(Y)=k_0^2e^{\lambda Y}$, holding the relation $Y=\ln(1/x)$, where $x$ is the Bjorken variable. The $x$-dependence of the saturation scale $Q_s^2\propto e^{\lambda Y}$ (with $\lambda \simeq 0.3$) is a general feature of the asymptotic solutions of the QCD nonlinear evolution equations at leading order.  We will follow the celebrated GBW model \cite{GolecBiernat:1998}, where $Q_s^2(Y)=Q_0^2e^{\lambda (Y-\tilde{Y}_0)}$ with $\tilde{Y}_0 = \ln (1/x_0)$ and $Q_0=1$ GeV, and use the fitted values of $x_0=4.2\times 10^{-5}$ and $\lambda=0.248$ from data \cite{Golec-Biernat:2017}.

The QCD dynamical entropy at rapidity $Y$, originating from $Y_0$, is defined as:
\begin{eqnarray}
\Sigma^{Y_0\rightarrow Y} & \equiv & \expval{\ln\left[\frac{P(k,Y)}{P(k,Y_0)}\right]}_Y \nonumber \\
& = & \int d^2k \, P(k,Y) \ln \left[\frac{P(k,Y)}{P(k,Y_0)}\right],
\label{dynamicalentropy01}
\end{eqnarray}
where $\expval{x(Y)}_Y$ denotes the average of the observable $x(Y)$ over the probability distribution in the final state at $Y$. The dynamical entropy measures the amount of disorder caused by the nonlinear rapidity evolution $Y_0 \rightarrow Y$. Thermodynamically, using the Jarzinsky-Crooks identity, the rapidity evolution can be understood as compression: initially, at $Y_0$, the initial color correlation size is $R_s(Y_0)$, and in the final stage, one observes $R_s(Y) < R_s(Y_0)$ for $Y_0 < Y$.

The transverse momentum distribution functions \( P(k,Y) \) are obtained from the UGD \( \varphi(k,Y) \) through the normalization procedure:
\begin{eqnarray}
P(Y, k) & = & \frac{1}{N} \, \varphi(Y,k), \label{kt-distr} \\
N & = & \int \varphi(Y,k) \, d^2k, \quad \int P(Y,k) \, d^2k = 1. \nonumber 
\label{Pdefinition}
\end{eqnarray}
The geometric scaling property simplifies the manipulation of integrals associated with the UGDs, where \( \varphi(k,Y) = \varphi(\tau) \), with \( \tau = k^2 / Q_s^2 \) as the scaling variable. In the macroscopic interpretation of QCD dynamical entropy, the transverse momentum serves as the phase space, while \( Y \) acts as the dynamic parameter in the Hatano-Sasa identity.

The total dynamical entropy density \( dS_D/dy \) is given by:
\begin{eqnarray}
\frac{dS_D}{dy} = C_m \mu \frac{R_h^2}{R_0^2} \, \Sigma^{Y_0\rightarrow Y},
\label{total-entropy}
\end{eqnarray}
where \( R_h \) is the size of the hadronic or nucleonic target, and \( R_0 = Q_s(Y_0) \). The parameter \( C_m = \frac{(N_c^2 - 1)}{4\pi N_c \alpha_s} \) represents the product of the color multiplicity \( (N_c^2 - 1) = 8 \) for the number of colors \( N_c = 3 \), multiplied by the typical gluon occupation number \( n_g \sim \frac{1}{4\pi N_c \alpha_s} \), with \( \alpha_s \approx \frac{1}{5} \) being the strong interaction coupling constant. The factor \( \mu \) denotes the average number of gluonic degrees of freedom within a transverse cell \( R_s(Y_0) \).

The unintegrated gluon distribution (UGD) is the main component for computing the dynamical entropy. In reference \cite{Peschanski:2013}, various Gaussian Color Glass Condensate (CGC) models were explored to quantitatively assess the dynamical entropy. The momentum probability distribution demonstrates geometric scaling, expressed as $P_{\mathrm{gauss}}(k,Y) = P_{\mathrm{gauss}}(\tau) \propto \Gamma^{-1}(\kappa) \tau^{\kappa-1} e^{-\tau}$. Here, $\kappa$ characterizes the low transverse momentum behavior of the dipole transverse momentum distribution (TMD), and for $\kappa = 2$, the gluon UGD exhibits the property of color transparency. The Gaussian UGD $\varphi_{\mathrm{GBW}}(\tau)$ can be evaluated using references \cite{GolecBiernat:1998, Kutak:2011} in the so-called GBW Gaussian model. In this framework, the saturation scale is associated with the peak gluon density, and the inclusive gluon production cross-section is calculated in the dilute-dense regime of proton-proton collisions. The GBW UGD is given by \cite{GolecBiernat:1998}:
\begin{eqnarray}
\varphi_{\mathrm{GBW}}(\tau) = C_{\mathrm{GBW}} \tau e^{-\tau/2},
\label{PCGCGAUS}
\end{eqnarray}
where $C_{\mathrm{GBW}} \equiv \frac{C_F \sigma_0}{8\pi^2 \alpha_s}$, with $C_F = \frac{(N_c^2 - 1)}{2N_c}$ being the Casimir constant and $\sigma_0 = 27.32 \,\text{mb}$ fitted from data \cite{Golec-Biernat:2017}. 

In the comparison to data, within the QCD color dipole picture, the relevant quantity is the color dipole cross section, $\sigma_{\mathrm{dip}}(\vec{r},x)$. It gives the probability of a color dipole of transverse size $\vec{r}$ at a given Bjorken $x$ to scatter with the (nucleon or nucleus) target. The UGD is related to the Fourier transform of the dipole cross section in the following way:
\begin{eqnarray}
\sigma_{\mathrm{dip}}(\vec{r},x) = \frac{4\pi^2}{3}\int \frac{dk^2}{k^2}\left[1-J_0(kr)  \right]   \alpha_s\varphi (x,k).
\end{eqnarray}

In the UGD GBW model, the dipole total cross-section $\sigma_{\mathrm{dip}}(r,Y)$ is given by \cite{GolecBiernat:1998}:
\begin{eqnarray}
\sigma_{\mathrm{dip}}(r,Y) = \sigma_0 \left(1 - e^{-r^2 Q_s^2 / 4}\right),
\label{gbwsigmadip}
\label{protondipole}
\end{eqnarray}
where $r$ denotes the color dipole size, characterizing the space configuration variable adjoint of the partonic transverse momentum $k$.

In reference \cite{Peschanski:2013}, using Gaussian CGC models, the author compared the formula (\ref{total-entropy}) with expression (25) in \cite{Kutak:2011}, where the saturation of the unintegrated gluon distribution (UGD) allows for the introduction of thermodynamic entropy with $Q_s^2(Y) = 2\pi T$, being $T$ the temperature. From this relation, the average number of gluonic degrees of freedom is identified as $\mu = \frac{3 \pi}{2}$, which matches both the aforementioned equations.

\begin{figure*}[t]
\centering
\begin{subfigure}{.5\textwidth}
  \centering
  \includegraphics[scale=0.2]{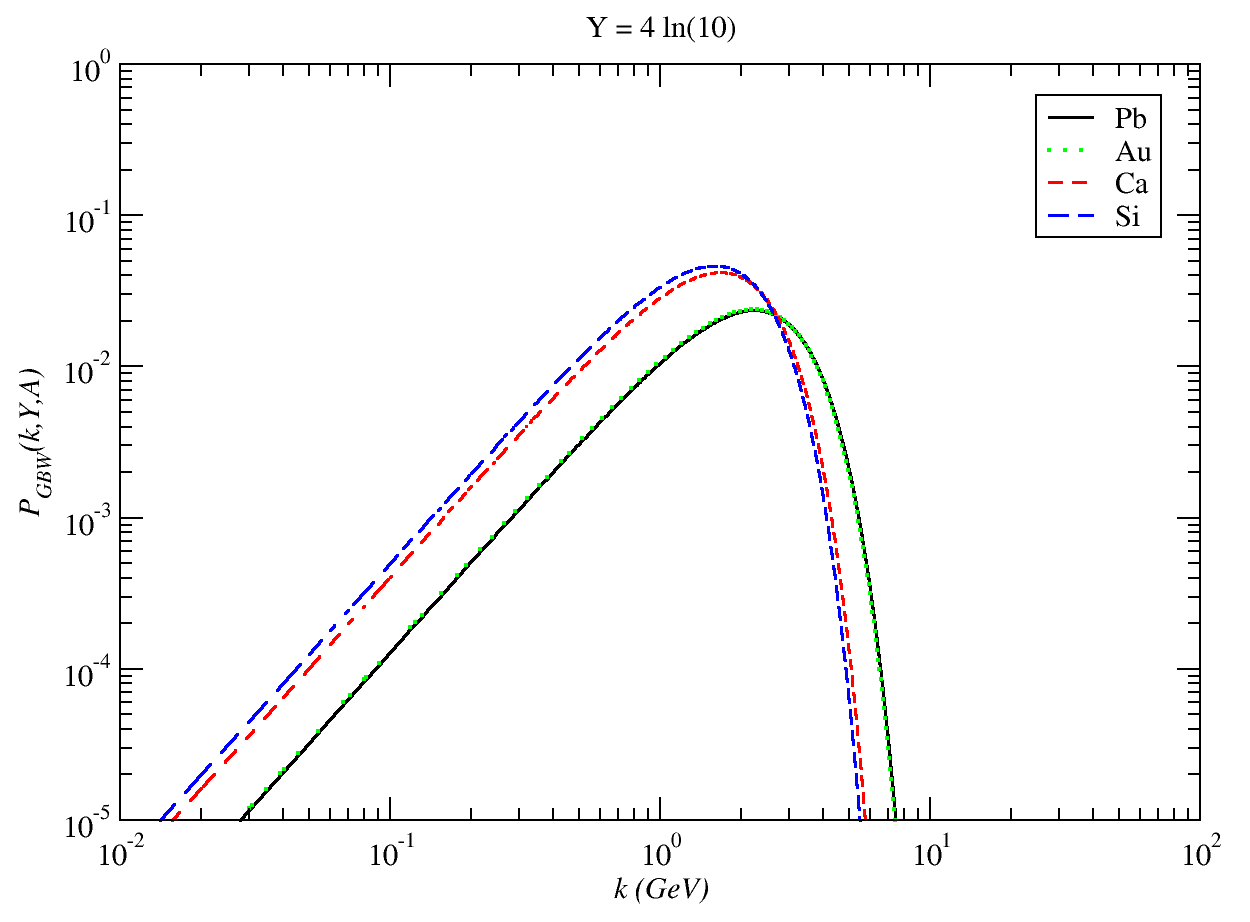}
  \caption{}
  \label{fig1:sub1}
\end{subfigure}%
\begin{subfigure}{.5\textwidth}
  \centering
  \includegraphics[scale=0.27]{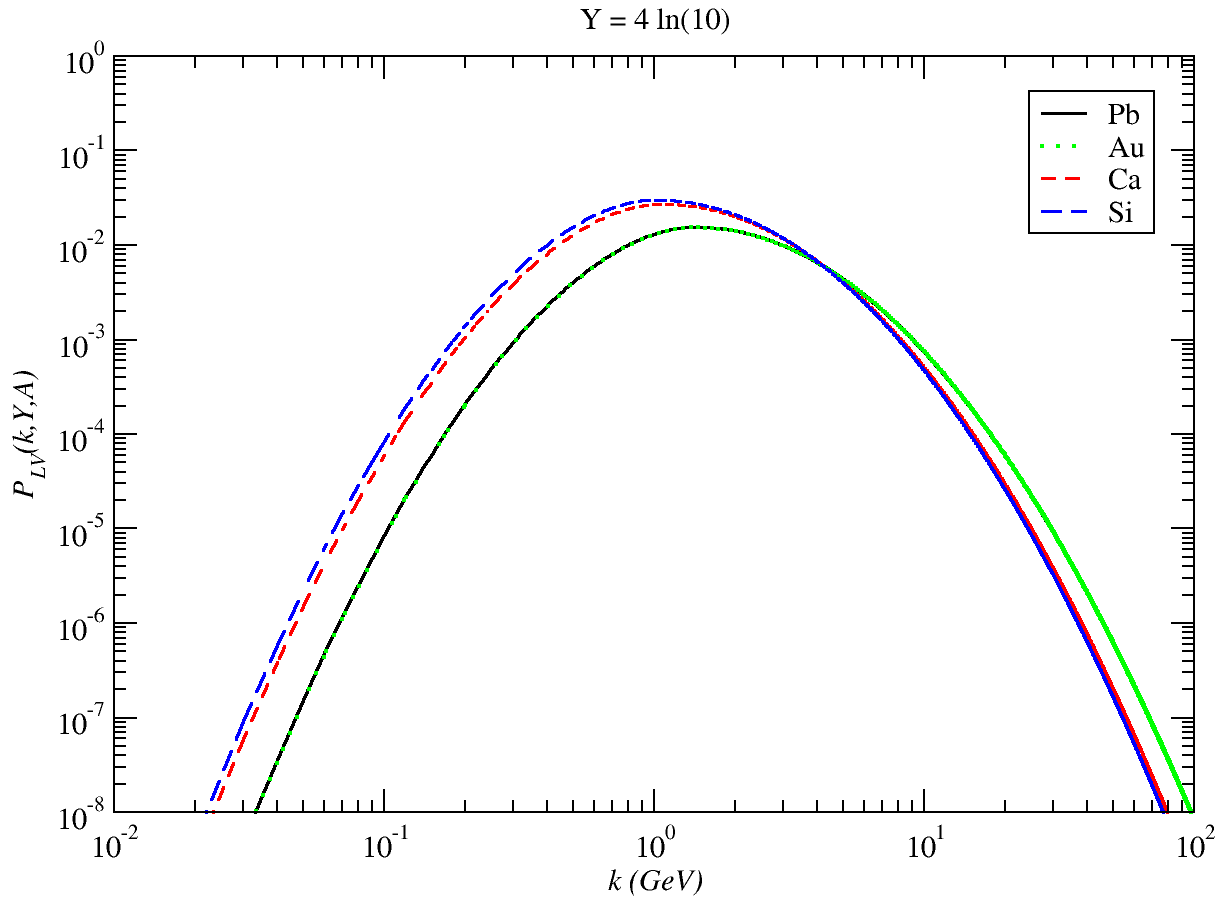}
  \caption{}
  \label{fig1:sub2}
\end{subfigure}
\caption{The nuclear transverse momentum probability, $P(Y,k)$, as a function of gluon transverse momentum $k$ for $Y = 4\ln 10$ ($x=10^{-4}$). Results for GBW model (left panel) and the LT one (right panel) are presented with both based on geometric scaling. Calculations are shown for lead (solid line), gold (dotted line), calcium (short dashed line) and silicon (long dashed line) nuclei.}
\label{fig:1}
\end{figure*}

Now, let us introduce another phenomenological model that incorporates geometric scaling. It is obtained by analyzing charged hadron production in proton-proton collisions by using a Tsallis-like distribution supplemented with the geometric scaling property. This model has been proposed in Refs. \cite{Moriggi:2020zbv,Moriggi:2020qla} (referred to as the MPM model), and the corresponding gluon unintegrated distribution  is given by
\begin{eqnarray}
\varphi_{\mathrm{MPM}}(\tau) = C_{\mathrm{MPM}} \frac{\tau(1 + a\tau^b)}{\left(1 + \tau\right)^{2 + a\tau^b}},
\label{FMPM}
\end{eqnarray}
where $C_{\mathrm{MPM}} \equiv \frac{3 \sigma_{0}}{4\pi^2 \alpha_{s}}$, and the values $a = 0.055$ and $b = 0.204$ are parameters of the model. Specifically for this UGD model, $\lambda = 0.33$ is used to fit the parameters $a$ and $b$. In the MPM UGD, the saturation scale has the same form as in the GBW UGD, $Q_s^2(Y) = Q_0^2e^{0.33 (Y-\tilde{Y}_0)}$.

To analyze a dipole UGD that incorporates more physical information and exhibits the correct theoretical behavior at both small and large transverse momenta, we will examine the one presented in Refs. \cite{Abir:2017mks,Siddiqah:2018qey}, called the LV model. This framework is derived as a general solution for $\varphi(k,Y)$, which accurately reflects both the McLerran-Venugopalan initial conditions and the Levin-Tuchin solution in their respective limits. It smoothly makes the transitions between these limits and provides a closer approximation to the numerical solution of the full leading-order Balitsky-Kovchegov equation, particularly in the deep saturation region. In this limit, the dipole gluon TMD approaches zero as $k\rightarrow0$, showing similarities with the Sudakov form factor \cite{Siddiqah:2018qey}. 

Starting with the dipole TMD at small transverse momentum from the Levin-Tuchin (LT) solution of the S-matrix, the corresponding UGD is given in the region $Q_s \gtrsim k \gtrsim \Lambda_{QCD}$:
\begin{eqnarray}
\varphi_{\mathrm{LT}}^{\mathrm{sat}}(\tau) = - C_{\mathrm{LV}} \ln \left(\frac{\tau}{4}  \right)\exp \left[-\varepsilon \ln^2 \left(\frac{\tau}{4}  \right)   \right].
\label{saturedLevin-Tuchin}
\end{eqnarray}
Here, $C_{\mathrm{LV}}\equiv \frac{N_c\sigma_0\varepsilon}{2\pi^3 \alpha_s}$. In the region of small transverse momentum, $\varphi_{\text{sat}}^{\text{LT}}$ is expressed as a series of Bell polynomials. The above equation represents the leading logarithmic approximation of the resummed series, with the constant $\epsilon \approx 0.2$ emerging from the saddle point condition along the saturation border. Just outside the saturation boundary, where $k \gtrsim Q_s$ but still near the saturation line, the QCD color dipole amplitude in transverse size space is approximated by $N(r, Y) \approx (r^2 Q_s^2)^{\gamma_s}$, where $\gamma_s \approx 0.63$ is the effective anomalous dimension near the saturation line. Under these conditions, the dipole TMD can be expressed as $\varphi_{\mathrm{LT}}^{\mathrm{dil}}(\tau) \propto C_{\mathrm{LV}} \tau^{-\gamma_s}$.

Imposing that the saturated and dilute distributions must be continuous, the UGD in the LV model is:
\begin{eqnarray}
\varphi_{\mathrm{LT}}(\tau)=\begin{cases}
-C_{\mathrm{LV}}\ln\left(\frac{\tau}{4}\right)e^{-\epsilon\ln^2(\tau/4)}, \quad \hbox{for}\quad \tau<1\\
C_{\mathrm{LV}}(d\tau)^{-\gamma_s}e^{-\epsilon\ln^2(\tau/4)}, \quad \hbox{for}\quad \tau<1
\end{cases}
\label{completeLevin-Tuchin}
\end{eqnarray}
where the parameter $d\approx0.5954$ assures the continuity of the different parts of this UGD.  

All these analytical phenomenological UGDs have been previously addressed in studies on QCD dynamical entropy for proton-proton collisions. Now, we will present two different strategies to extend the dynamical entropy studies to proton-nucleus collisions. The first strategy involves utilizing the geometrical scaling properties, while the second strategy focuses on adapting the dipole-proton cross section from the proton-proton case to the proton-nucleus scenario using the Glauber-Gribov framework.

\begin{figure*}[t]
\centering
\begin{subfigure}{.5\textwidth}
  \centering
  \includegraphics[scale=0.2]{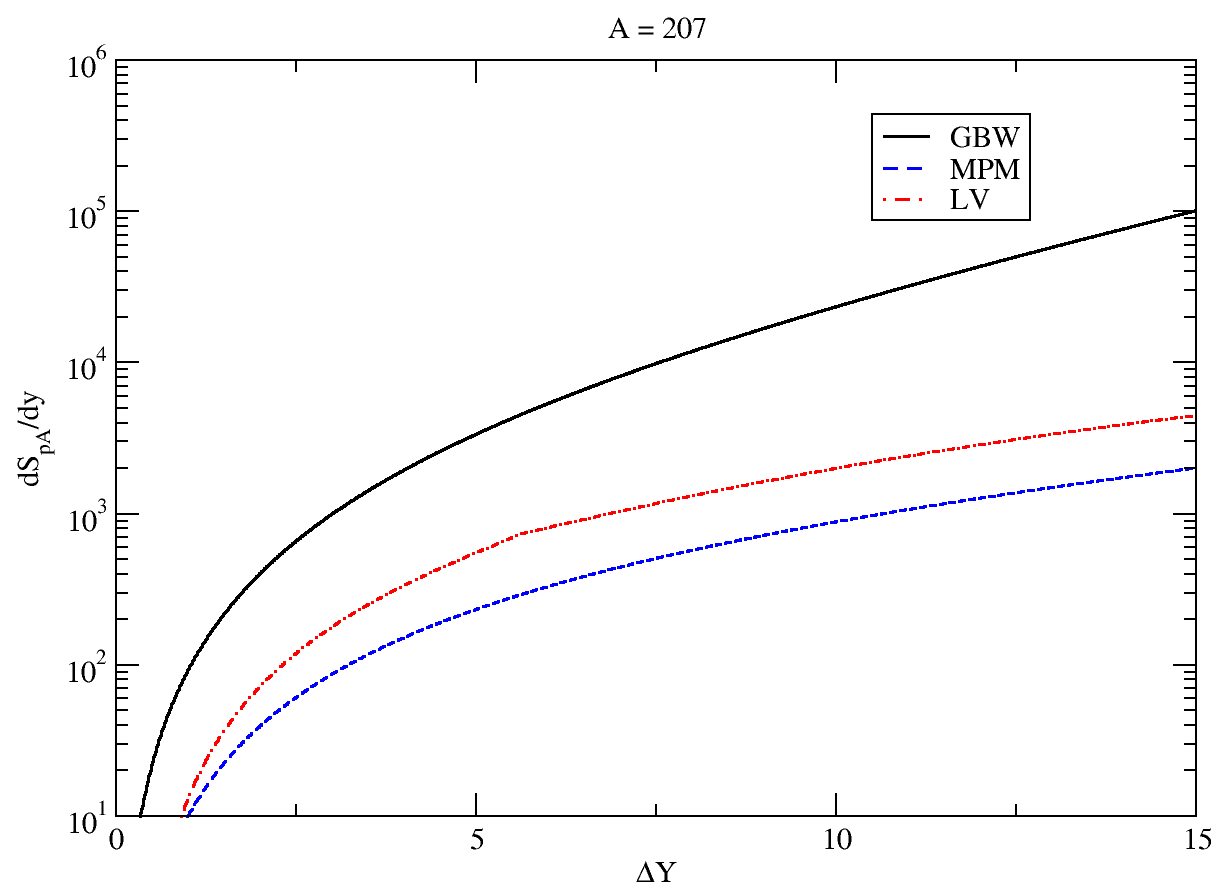}
  \caption{}
  \label{fig2:sub1}
\end{subfigure}%
\begin{subfigure}{.5\textwidth}
  \centering
  \includegraphics[scale=0.2]{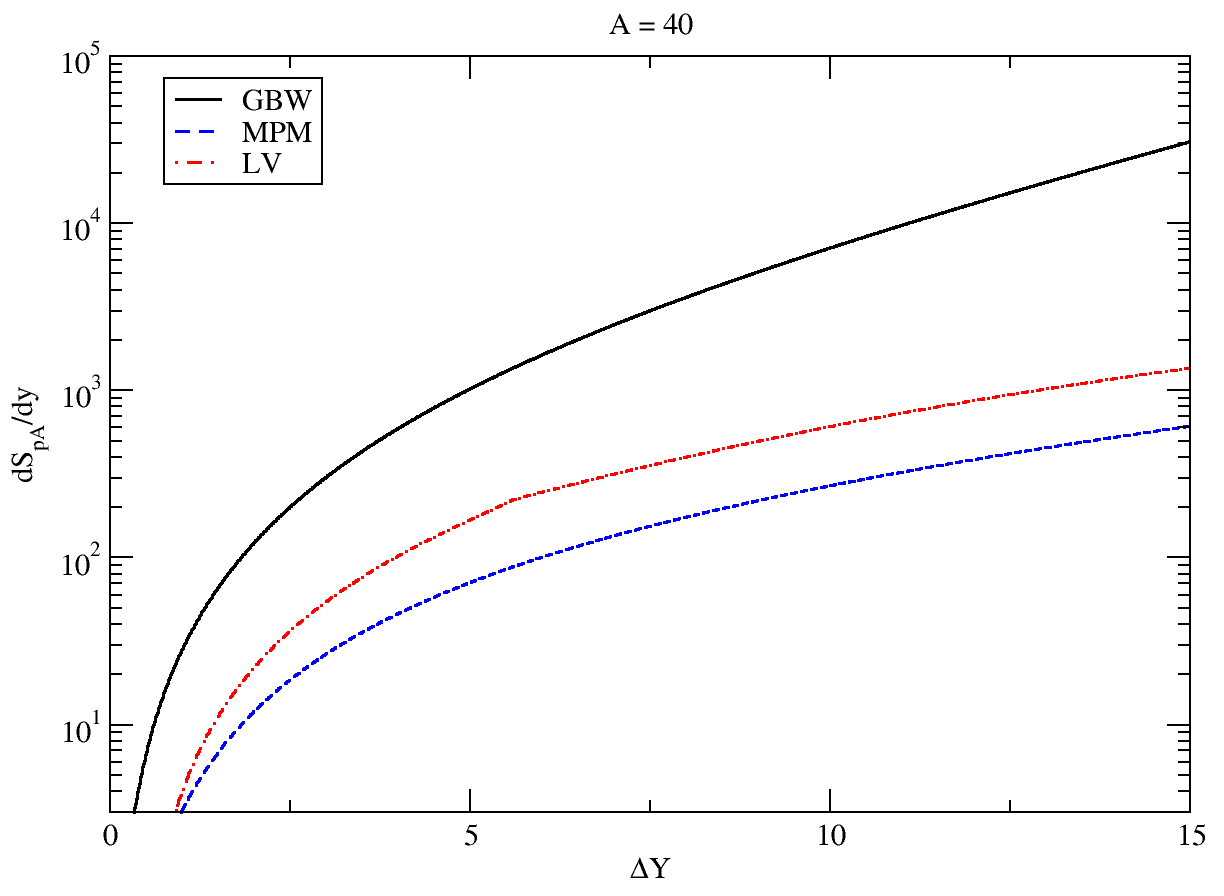}
  \caption{}
  \label{fig2:sub2}
\end{subfigure}
\caption{Total QCD dynamical entropy $dS_D/dy$ produced in a $pA$ collision as a function of $\Delta Y=Y-Y_0$ with $Y\approx 4.6$ for the GBW (solid lines), MPM (dashed lines) and LV (dotted-dashed lines) UGD models for lead (left) and calcium (right).}
\label{fig:2}
\end{figure*}

\subsection*{Strategy I: Geometric Scaling Procedure}
\label{subsection1sec2}

If one wants to estimate proton-nucleus collision observables, the geometrical scaling property can be applied, where the ratio between the cross-section of the virtual photon-nucleus and the transverse area of the target can be absorbed by the saturation scale dependent on the atomic mass number \( A \), i.e., \( \sigma^{\gamma^*A}(\tau_A)/\pi R_A^2=\sigma^{\gamma^*p}(\tau)/\pi R_p^2 \), with \( R_A=(1.12A^{1/3}-0.86A^{-1/3}) \) fm being the nuclear radius \cite{Armesto:2004}. Thus, it is necessary to adapt the transverse cross-section \( \sigma_0 \rightarrow \sigma_A \) and the saturation scale \( Q_s^2(Y) \rightarrow Q_{s,A}^2(Y) \). Specifically, for the nuclear saturation scale \( Q_{s,A}(Y) \):
\begin{eqnarray}
Q_{s,A}^2(Y)=\left(\frac{R_p^2A}{R_A^2}\right)^\Delta Q_s^2(Y),
\label{gsprocedure01}
\end{eqnarray}
where \( \Delta \approx 1.27 \) and \( R_p \approx 3.56 \) GeV \cite{Armesto:2004}.

Now, in the dynamical entropy computation, the normalization procedure (\ref{Pdefinition}) is identical to the proton case as shown in our previous results \cite{Ramos2022}, where all dependencies on the cross-section normalization \( \sigma_0 \) are negligible due to model-specific proportional constants \( C_\mathrm{GBW} \), \( C_\mathrm{MPM} \), and \( C_\mathrm{LV} \), in that, the operation cuts off its dependency. Therefore, by doing the replacement \( Q_s^2(Y) \rightarrow Q_{s,A}^2(Y) \), the transverse momentum probability distributions for the GBW, MPM, and LV models in the proton-nucleus case are:

\begin{eqnarray}
P_{\mathrm{GBW}}^A (\tau_A)=\frac{\tau_A e^{-\tau_A/2}}{4\pi Q_{s,A}^2},
\label{Pgbw}
\end{eqnarray}

\begin{eqnarray}
P_{\mathrm{MPM}}^A(\tau_A)=\frac{1}{\pi \xi Q_{s,A}^2}\frac{\tau_A(1+a\tau_A^b)}{(1+\tau_A)^{2+a\tau_A^b}},
\label{Pmpm}
\end{eqnarray}

\begin{eqnarray}
P_{\mathrm{LV}}^A(\tau_A)=\begin{cases}
-\frac{\ln\left(\frac{\tau_A}{4}\right)}{8\pi Q_{s,A}^2} e^{-\epsilon \ln^2\left(\frac{\tau_A}{4}\right)}, & \text{for } \tau_A < 1; \\
\frac{(d\tau_A)^{-\gamma_s}}{8\pi Q_{s,A}^2} e^{-\epsilon \ln^2\left(\frac{\tau_A}{4}\right)}, & \text{for } \tau_A \geq 1.
\end{cases}
\label{Plv}
\end{eqnarray}

In these equations, the scaling variable is now \( \tau_A = k^2/Q_{s,A}^2 \), and \( \xi = 4.346 \) is the normalization factor for the MPM model.

\subsection*{Strategy II: Glauber Gribov Formalism}
\label{subsection2sec2}

In the Glauber-Gribov framework, to obtain a nuclear UGD, the  dipole-proton total cross-section $\sigma_{\mathrm{dip}}(r,Y)$ is replaced by the nuclear one, $\sigma_{dA}(x,r)=\int d^2b \, \sigma_{dA}(x,r,b)$, with:
\begin{eqnarray}
\sigma_{\mathrm{dA}}(Y,r,b)=2\left[1-\exp\left(-\frac{1}{2}T_A(b)\sigma_{\mathrm{dip}}(Y,r)\right)\right].
\label{glaubergribov01}
\end{eqnarray}
In Eq. (\ref{glaubergribov01}), $T_A(b)$ is the thickness function, the nuclear profile function $T_A=\int_{-\infty}^{+\infty}\rho_A(z,\vec{b})$, normalized to the atomic mass number, $\int d^2b T_A(b)=A$. In this work, we used the Woods-Saxon parametrization for the nuclear density $\rho_A$.

The nuclear UGD is given by the expression:
\begin{eqnarray}
\varphi_A(Y,k)=-\frac{N_c k^2}{4\pi^2\alpha_s}\int \frac{d^2bd^2r}{2\pi}e^{i\vec{k}\cdot \vec{r}}\sigma_{\mathrm{dA}}(Y,r,b)
\label{glaubergribov02}
\end{eqnarray}
In particular, for the GBW model in the small-$x$ regime, one can use the  dipole-proton cross-section (\ref{protondipole}) and the nuclear UGD is \cite{Armesto:2002}:
\begin{eqnarray}
\varphi_A^{GBW}(x,k)&=&\frac{N_c}{\pi^2 \alpha_s}\frac{k^2}{Q_s^2}\int d^2b\sum_{n=1}^{\infty}\frac{(-B)^n}{n!}\nonumber\\
&\times & \sum_{\ell=1}^n C_{\ell}^n\frac{(-1)^{\ell}}{\ell}e^{-k^2/\ell Q_s^2},
\label{glaubergribov03}
\end{eqnarray}
where $C_{\ell}^n$ is the combination formula and $B=\frac{1}{2}T_A(b)\sigma_0$.

Using strategies I and II, one can compute the dynamical entropy. In  the next section, the results will be evaluated from these procedures in order to obtain the nuclear UGD and its consequences.
 
\begin{figure}[t]
\includegraphics[scale=0.25]{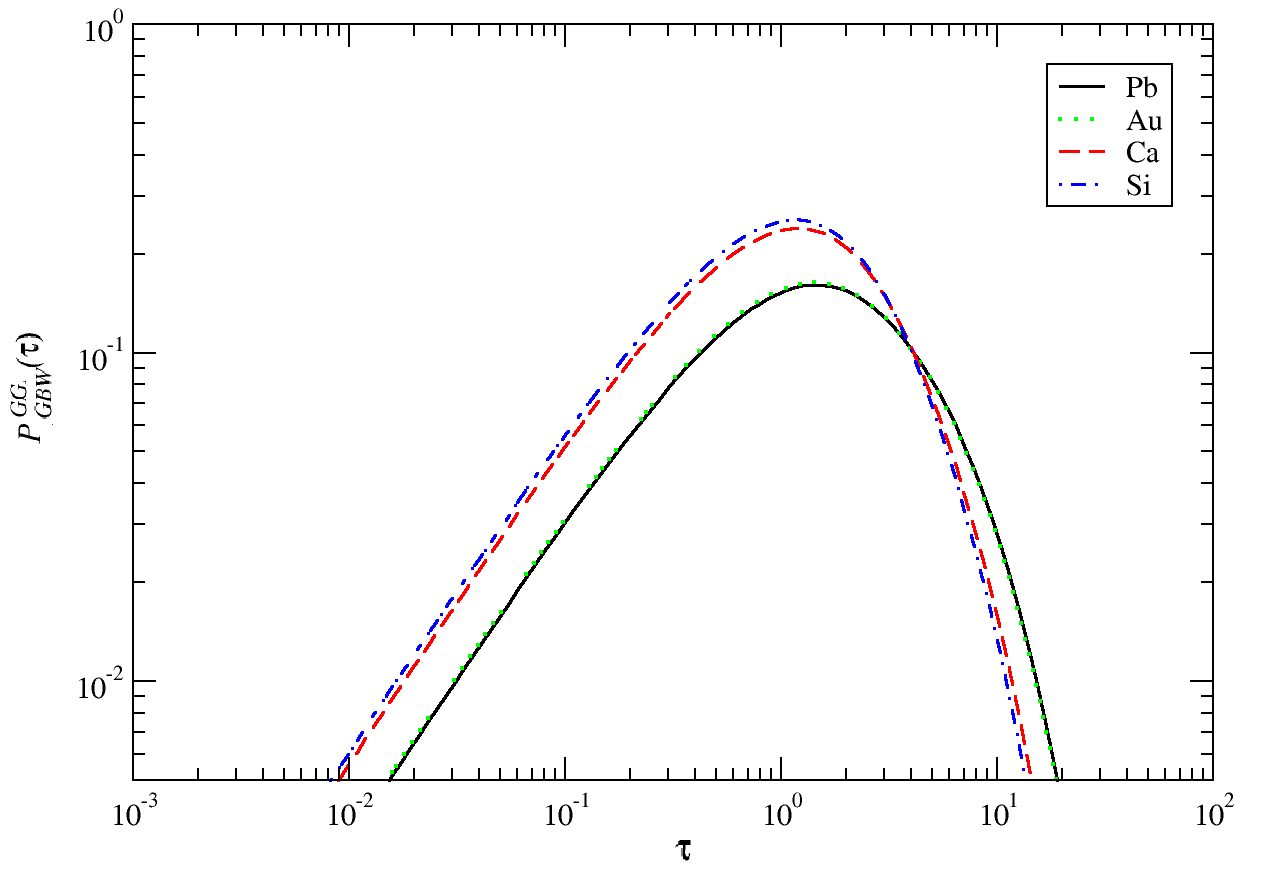}
\caption{Transverse momentum probability distributions  from the Glauber-Gribov framework as function of the nuclear scaling variable $\tau_A=k^2/Q_{s,A}^2$. In the figure,  results are presented for four nuclei: lead (solid line), gold (dotted line), calcium (short dashed line) and silicon (long dashed line).}
\label{fig:3}
\end{figure}

\section{Results and discussions}
\label{sec3}

\begin{figure}[t]
\includegraphics[scale=0.25]{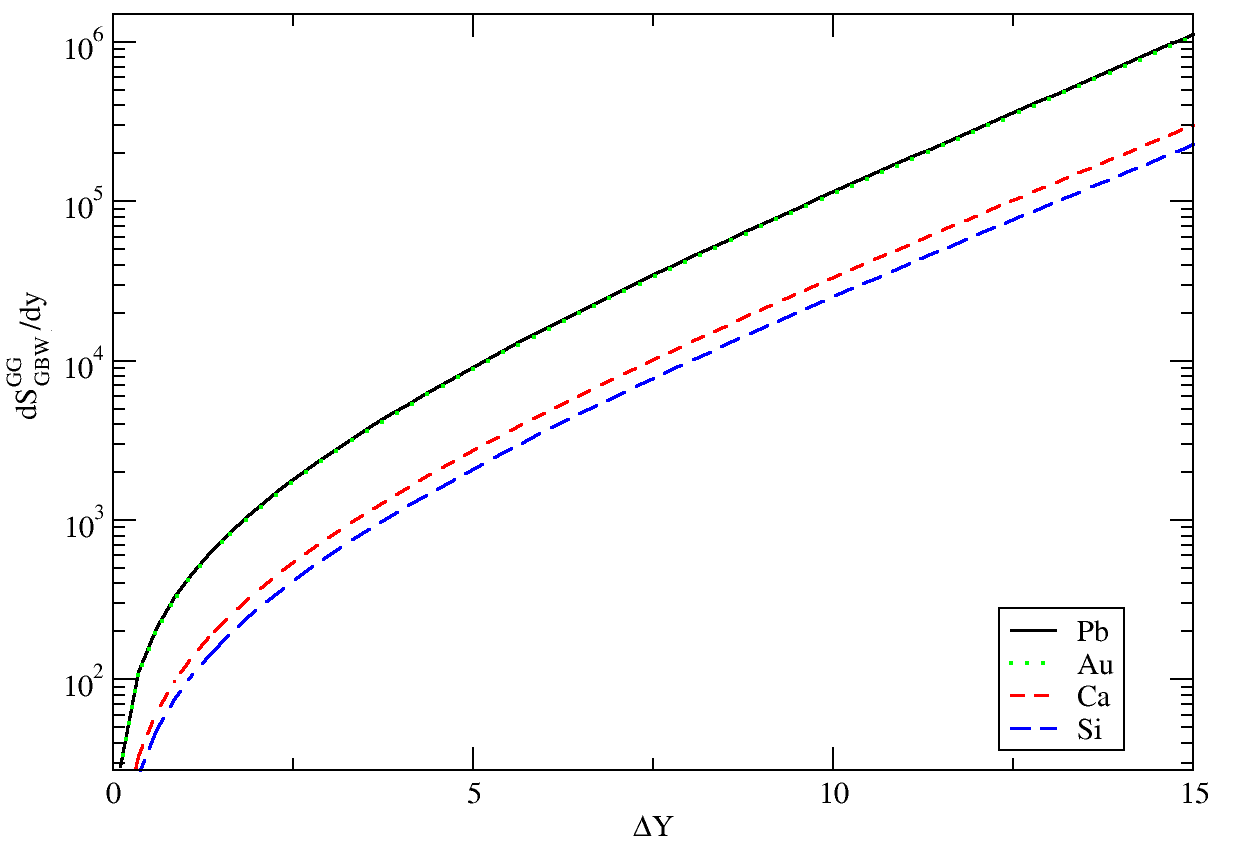}
\caption{Total dynamical entropy in proton-nucleus collisions corresponding to the QCD evolution in rapidity, $Y_0\rightarrow Y$, within the range $\Delta Y=[0,15]$ in the Glauber-Gribov framework. Same notation as previous figure.}
\label{fig:4}
\end{figure}
\begin{figure*}[t]
\includegraphics[scale=0.4]{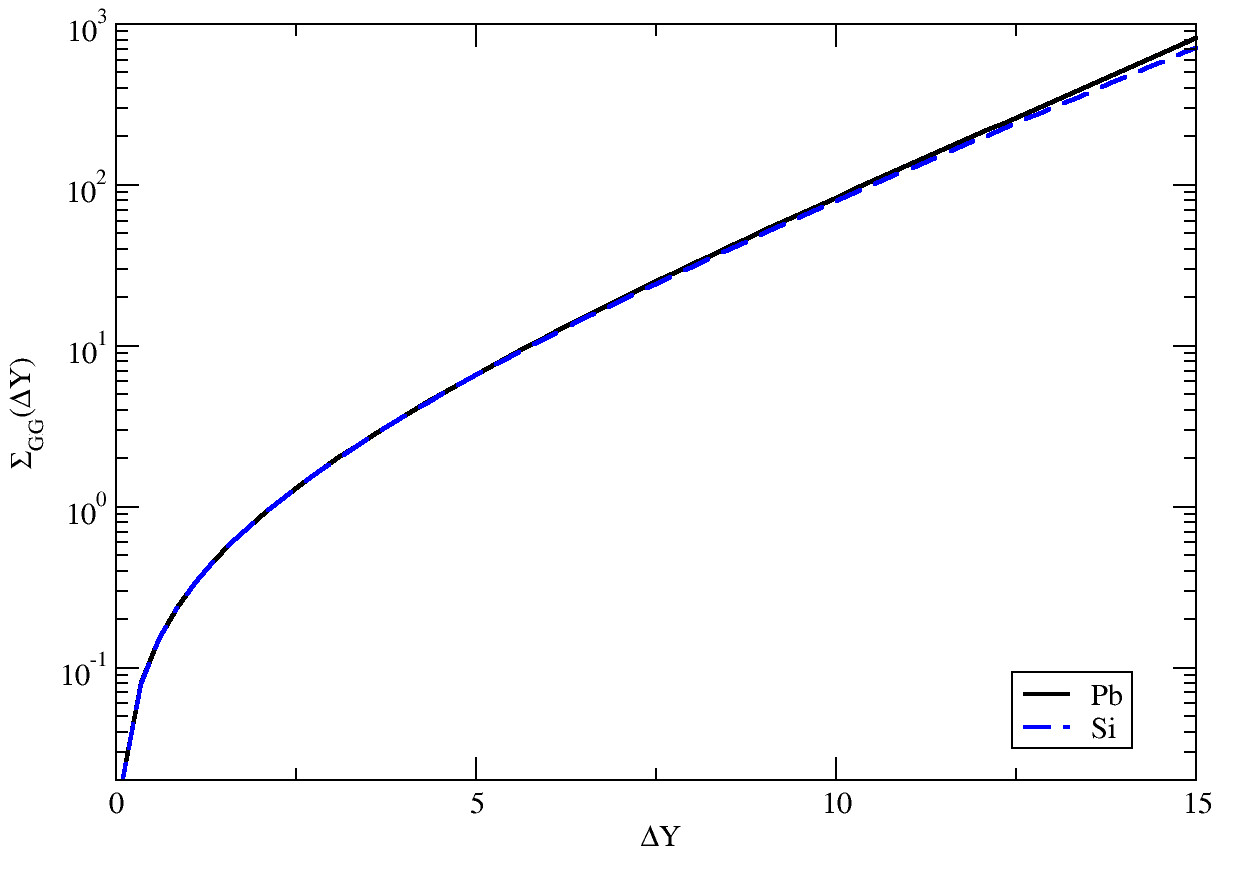}
\caption{Nuclear QCD dynamical entropy in proton-nucleus collisions corresponding to the QCD evolution in rapidity, $Y_0\rightarrow Y$, within the range $\Delta Y=[0,15]$ in the Glauber-Gribov approach. The entropy is computed for lead (solid line) and silicon (dashed lines).}
\label{fig:5}
\end{figure*}

In Fig. \ref{fig:1}, the nuclear transverse momentum probability distribution is plotted as a function of transverse momentum for the UGD from GBW (\ref{fig1:sub1}) and LV (\ref{fig1:sub2}) models which present geometric scaling. The rapidity is fixed at $Y=4\ln(10)$ for some representative nuclei: lead (solid line), gold (dotted line), calcium (short dashed line), and silicon (long dashed line). The peak in the transverse momentum occurs at different values for each model or nucleus. In the case of the GBW model, $k_{\mathrm{max}}=\sqrt{2}Q_{s,A}$ and for the LV UGD, one has $k_{\mathrm{max}}=e^{\gamma_s/4\epsilon}Q_{s,A}\approx 2.2 Q_{s,A}$. For heavy nuclei, the peak is lower for the smaller ones because, in the geometric scaling framework, $k_{\mathrm{max}}\sim 1/R_A^{2\Delta}$ for both models.

For the dynamical entropy in the geometric scaling strategy, using the nuclear transverse momentum probability distributions (\ref{Pgbw})-(\ref{Plv}), we obtained the same result as shown in Fig. 2 of our previous work \cite{Ramos2022}. To understand this, we can evaluate the dynamical entropy in the following form:
\begin{eqnarray}
\Sigma^{Y_0\rightarrow Y}=\pi Q_{s,A}^2(Y)\int_0^{\infty}d\tau_A P(\tau_A)\ln\left[\frac{P(\tau_A)}{P(\tau_{A}^0)}\right],
\label{dynamicalentropy02}
\end{eqnarray}
with $\tau_{A}^0=k^2/Q_{s,A}^2(Y_0)$. From this, $k^2=\tau_AQ_{s,A}^2(Y)=\tau_A^0Q_{s,A}^2(Y_0)$, and it is useful to define the ratio,
\begin{eqnarray}
\frac{Q_{s,A}^2(Y)}{Q_{s,A}^2(Y_0)}=\frac{\left(\frac{R_p^2A}{R_A^2}\right)^{\Delta}Q_s^2(Y)}{\left(\frac{R_p^2A}{R_A^2}\right)^{\Delta}Q_s^2(Y_0)}=e^{\lambda\Delta Y}\equiv s,
\label{sdefinition}
\end{eqnarray}
where $\Delta Y=Y-Y_0$ with $Y_0\approx 4.6$ ($x_0=0.01$). We consider this initial rapidity because the values of $x \leq x_0$ correspond to the limit of validity for the application of the phenomenological UGD models considered here. Thus, partons  initially populate a transverse area proportional to the initial color correlation size $R_0(Y_0) = 1/Q_s(Y_0)$.

From expression (\ref{dynamicalentropy02}), the ratio between \( P(\tau_A) \) and \( P(s\tau_A) \) can be analyzed for the different expressions of the nuclear transverse momentum probability distributions in the geometric scaling adapting strategy (\ref{Pgbw})-(\ref{Plv}). For the GBW model, one can see:
\begin{eqnarray}
\ln\left[\frac{P_{\mathrm{GBW}}(\tau_A)}{P_{\mathrm{GBW}}(s\tau_A)}\right]=2\lambda\Delta Y+k^2[R_s^2(Y_0)-R_s^2(Y)].
\label{ratioGBW}
\end{eqnarray}
Plugging this result into expression (\ref{dynamicalentropy02}), one gets an equivalent expression as obtained in Eq. (14) of Ref. \cite{Peschanski:2013}  and it recovers the proton dynamical entropy:
\begin{eqnarray}
\Sigma_{\mathrm{GBW}}^{Y_0\rightarrow Y}(\Delta Y)=2\left(e^{\lambda \Delta Y}-1-\lambda\Delta Y\right),
\label{protongbwdynamicalentropy}
\end{eqnarray}

A similar procedure can be done for the dynamical entropy of the MPM model,  \( \Sigma_{\mathrm{MPM}}^{Y_0\rightarrow Y} \), once it presents geometric scaling property:
\begin{eqnarray}
\Sigma^{Y_0\rightarrow Y}_{\mathrm{MPM}}&=&\frac{1}{\xi }\int_0^{\infty}d\tau_A\frac{\tau_A(1+a\tau_A^b)}{(1+\tau_A)^{2+a\tau_A^b}}\nonumber\\
&\times&\ln\left[\frac{(1+a\tau_A^b)(1+s\tau_A)^{2+as^b\tau_A^b}}{s^2(1+as^b\tau_A^b)(1+\tau_A)^{2+a\tau_A^b}}\right].
\label{protonMPMdynamicalentropy}
\end{eqnarray}
This expression also returns the proton dynamical entropy shown in Fig. 2 of Ref. \cite{Ramos2022}. The normalization procedure (\ref{Pdefinition}) removes all nuclear dependence on the target transverse size, $S_{\perp}^{A} = \pi R_A^{2}$. Moreover,  we have shown that the relation in Eq. (\ref{sdefinition}) makes the nuclear result identical to the proton one. The same effect can be shown for the LV UGD case, for both diluted and saturated contributions.

In Fig. \ref{fig:2} the entropy density (\ref{total-entropy}) is computed for lead (left panel) and for calcium (right panel) for all models based on the geometric scaling phenomenon: GBW (solid line), MPM (dashed line), and LV (dash-dotted line) in the range $\Delta Y=[0,15]$. Although the nuclear dynamical entropy is independent of $A$, its density is related to the size of the nuclear radius as $\frac{d S_D}{dy} \sim R_A^2$. In the definition proposed in Ref. \cite{Peschanski:2013}, one must take into account the ratio between all available unit cells in the CGC medium, $\sim \frac{\pi R_A^2}{\pi R_0^2}$.

Now we will discuss our results for nuclear dynamical entropy obtained via the Glauber-Gribov formalism. In Fig. \ref{fig:3}, the nuclear transverse momentum probability distribution is shown as a function of the scale variable $\tau_A=k^2/Q_{s,A}^2$. Similarly to the geometric scaling strategy, we computed these distributions for the same nuclei (and labels) as in Fig. \ref{fig:1}. By using strategy II, considering the Glauber-Gribov approach, the dynamical entropy shows $A$-independence in Fig. \ref{fig:4}, also plotted as a function of $\Delta Y$ in the interval $[0,15]$. Although obtaining the UGD involves a more complex process via eq. (\ref{glaubergribov03}), it appears that geometric scaling and the normalization procedure also eliminate the $A$ dependence in the dynamical entropy. A notable difference compared to strategy I is that this result does not reduce to the proton case, and it is substantially larger.
 
Our last result is the QCD dynamical entropy density for proton-nucleus collisions shown in Fig. \ref{fig:4}. In this case, we observe a difference in the total entropy generated in proton-nucleus collisions because, as in the geometric scaling case, we have the dependency $\frac{d S_D}{dy}\sim R_A^2$. The results are plotted as a function of $\Delta Y$ in the same interval and for the same nuclei as in the other figures. We call attention to the fact that the glasma phase of the QGP is not being discussed here. The extraction of the gluon distribution in the pre-equilibrium stage is a complex task.  It is expected that the geometric scaling is still valid in this situation and the dynamical entropy density of the glasma state could be a candidate for the initial entropy density in the context of the strong coupling thermalization mechanism (see Ref. \cite{Peschanski:2013} for a discussion about this point).

Finally, we compare the QCD dynamical entropy discussed above to other formalisms for the entropy of partons at high energies. One of them is the  entanglement (von Neumann) entropy evaluated in the Color Glass Condensate (CGC) approach, which is  obtained by taking into account  soft gluons in the wavefunction of a fast-moving hadron. It is obtained by using  the reduced density matrix \cite{Kovner:2015hga} for these soft modes in the McLerran-Venugopalan (MV) model. The calculation can be done in the field basis or in the number representation basis \cite{Duan:2020jkz}. The leading contribution in terms of the saturation scaling has the following form \cite{Duan:2020jkz}:
\begin{eqnarray}
\label{CGCSEE}
S_{EE}^{\mathrm{CGC}}\propto \frac{1}{2}S_{\perp}\frac{C_F}{4\pi}\tilde{Q}_s^2,
\end{eqnarray}
where $\tilde{Q}_s^2(x)=(9/4)(x_0/x)^{\lambda}$ and $S_{\perp}$ are the gluon saturation scale and the transverse size of the hadron/nucleus, respectively. 

Another formalism  is the Wehrl entropy in QCD \cite{Hagiwara:2017uaz}, which is the semiclassical analogue of the von Neumann entropy. It is obtained in terms of quantum phase space distributions, like Wigner or Husimi ones.  The QCD Wigner distribution, $W (x,\vec{k}),\vec{b})$, is a generalization of the usual collinear parton distribution functions and depends on parton transverse momentum, $\vec{k}$,  impact parameter, $\vec{b}$, and longitudinal momentum fraction, $x$. In case the Wigner distribution is positive definite, then the Wehrl entropy associated with the gluons can be defined in the following way \cite{Hatta:2016dxp},
\begin{eqnarray}
S_W &= &  -\int d^2bd^2k\, xW_{g}(x,k,b)\ln \left[ xW_{g}(x,k,b) \right], 
\label{WehrlS}
\end{eqnarray}
where $xf_g(x)=xg(x)$ is the usual collinear gluon distribution, with $xf_{g}(x) =  \int d^2bd^2k \,xW_{g}(x,k,b)$.

For sake of illustration, we will consider the Weiszacker-Williams (WW) gluon Wigner distribution, which can be computed in a quasi-classical approximation \cite{Kovchegov:1998bi,Dominguez:2011wm}. It is written in terms of the forward  $S$-matrix of a QCD color dipole of transverse size $\vec{r}$, transverse momentum $\vec{k}$ at impact parameter $\vec{b}$ in the adjoint representation, ${\cal{S}}_A$,
\begin{eqnarray}
xW_g(x,k,b) = \frac{C_F}{2\pi^4 \alpha_s}\int d^2\vec{r} \,\frac{e^{i\vec{r}\cdot \vec{k}}}{r^2}\left( 1-{\cal{S}}_A(x,\vec{r},\vec{b}) \right). \nonumber\\
\end{eqnarray}

The WW Wigner distribution can be analytically evaluated in the case of a Gaussian form for the S-matrix, ${\cal{S}}_A(x,r,b)=\exp [-\vec{r}^2\tilde{Q}_s^2(x,b)/4]$, where $\tilde{Q}_s^2(x,b) =(N_c/C_F)Q_s^2(x,b)$ is the impact parameter dependent  gluon saturation scale. Specifically, for the Gaussian $S$-matrix one obtains,
\begin{eqnarray}
xW_g(x,k,b) = \frac{C_F}{2\pi^3\alpha_s}\Gamma\left( 0,\frac{k^2}{\tilde{Q}_s^2(x,b)}\right),
\label{WW}
\end{eqnarray}
which is positive definite with $\Gamma$ being the incomplete gamma function. In Ref. \cite{Ramos:2020kaj} we have shown that by using the impact-parameter(quark)  saturation scale from the b-CGC model \cite{Rezaeian:2013tka}, where $Q_s^2(x,b)=(x_0/x)^{\lambda}\exp [ -b^2/2\gamma_s B_{\mathrm{CGC}} ]$, one obtains the following approximation for the Wehrl entropy,
\begin{eqnarray}
\label{SWap}
S_W  \approx  - \frac{2F\gamma_s B_{\mathrm{CGC}}N_c}{2\pi \alpha_s}\,Q_s^2(x)= -\frac{2FN_c S_{\perp}}{6\pi^2\alpha_s}\,Q_s^2(x),
\end{eqnarray}
where the quantity $B_G=\gamma_s B_{\mathrm{CGC}}$ is related to the electromagnetic proton radius. From the equation above, the parametric behavior is $S_W\propto S_{\perp}Q_s^2(x)$. This behavior is quite similar to the von Neumann entropy computed in the CGC formalism. Notice that for realistic impact parameter-dependent saturation scale, $Q_s(x,b)$, the dipole Wigner distribution is not positive definite \cite{Hagiwara:2017uaz}.

Based on the geometric scaling arguments, where the following replacement is done $S_{\perp}^h\rightarrow S_{\perp}^A =\pi R_A^2$ and $Q_{s,p}\rightarrow Q_{s,A}^2\propto A^{1/3}Q_{s,p}^2$, both the nuclear Wehrl and CGC entropies behave like $S^A\sim A Q_{s,p}^2(x)=Ae^{\lambda Y}$. This means that in the nuclear case, these entropies are additive with respect to the hadron ones. This feature is consistent with the entropy being an extensive variable. On the other hand, we have shown that the QCD dynamical entropy does not present this property. As a last comment, the parton entanglement entropy proposed in Ref. \cite{Kharzeev:2017qzs} does not present the extensibility property. For the relevant kinematic region investigated here, where the average transverse momentum of partons is proportional to the saturation scale, the entanglement entropy can be written in the form \cite{Ramos:2020kaj}:
\begin{eqnarray}
 \label{SEESAT}
  S_{EE}^{KL}= \ln \left[Q_{s,A}^2(x)\right]+ S_0,\quad Q_{s,A}^2\sim A^{1/3}Q_{s,p}^2,
 \end{eqnarray}
where $S_0 = \ln[3(e-2)R_A^2/4e\pi \alpha_s ]$. The inclusion of the nuclear effects in this simplified analysis gives an overall additive correction of order $\ln (A)$ to the entanglement entropy for a nucleus compared to the hadronic one (for lead, $\ln (A)\simeq 5$).

As a final remark, the QCD dynamical entropy can be faced as a relative entropy. In general, other entropic quantities
can be regarded as special cases of the relative entropy. As discussed in the introduction section, relative entropy has the advantage that it generalizes to continuous probability distributions.  In the continuum limit, $S(P\parallel  Q) = \int du P(u)[\ln P(u)-ln Q(u)]$, where $P(u)$ and $Q(u)$ are probability densities. In our case, $u\rightarrow \vec{k}$ (with $du\rightarrow d^2k$), $P(u) = P(\vec{k},Y)$, and $Q(u) = P(\vec{k},Y_0)$. This means that the distribution $Q$ is related to the $P$ one by the QCD small-$x$ parton evolution. In the geometric scaling framework, the $Q$ distribution is obtained by just rescaling the continuum variable $\tau_A\rightarrow s\tau_A $, with $s=e^{\lambda \Delta Y}$, in the $P$ distribution (see Eqs. (17-20)). 
The relative entropy is positive and one has $S(P\parallel Q) = 0$ if and only if $P(u) = Q(u)$. It is not a distance measure in the mathematical sense due to its asymmetry, $S(P\parallel Q)\neq S(Q\parallel P)$. However, in the limit where $P$ and $Q$ are close to each other, the relative entropy satisfies the properties of a metric \cite{Schrofl:2023hnz}. Let us  consider a set of probability distributions $P(\tau)(u)$ where $\tau$ is a multi-dimensional parameter. Near to some point $\tau_0$ an expansion can be made, $S(P(\tau)\parallel P(\tau_0))=1/2(\tau^j-\tau_0^j)(\tau^k-\tau_0^k)g_{jk}(\tau_0)+\ldots $, where the  constant and linear terms vanish because  $S(P\parallel Q)$ is positive semi-definite and vanishes at $P=Q$. The quantity $g_{jk}(\tau_0) $ is known as the Fisher information metric (FIM) or Fisher-Rao metric \cite{Nielsen2023}. The FIM is symmetric and plays the role of a Euclidean, positive semi-definite metric on the space of parameters of probability distributions $P$.

Therefore, this sort of relative entropy cannot be directly compared to the measured hadron entropy like the other notions of entropy discussed before. The phenomenology of relative entropy applied to high-energy collisions is still incipient (see Ref. \cite{Bose:2020shm} for recent applications to 2-2 scattering) and it is out of the scope of the present study. In practical applications to heavy ion collisions, it seems promising the recent work \cite{Dowling:2020nxc} on the discussions about the second law of thermodynamics for relativistic fluids formulated with relative entropy. 

Concerning its quantum mechanics version, the quantum relative entropy can be faced as a limit of the quantum relative Rényi entropy. The quantum  Rényi entropy, $S_N(\rho)$, and the quantum Rényi
 relative entropy, $S_N(\rho\parallel \sigma)$, for a parameter $N > 0$ are given by, 
\begin{eqnarray}
S_N(\rho) &=&    \frac{1}{1-N}\ln \mathrm{Tr}[\rho^N],\\
S_N(\rho\parallel \sigma) &=& \frac{1}{1-N}\ln \frac{\mathrm{Tr}[\rho \sigma^{N-1}]}{\mathrm{Tr}[\rho^N]} \nonumber \\
&=& \frac{1}{1-N}\ln \mathrm{Tr}[\rho \sigma^{N-1}] - S_N(\rho).
\end{eqnarray}
In the limit $N\rightarrow 1$ the quantum Rényi relative entropy approaches the quantum relative entropy. As well known, in this very same limit the quantum Rényi entropy becomes von Neumann's entropy. 

From the practical point of view, as advocated in Ref. \cite{Peschanski:2013}, the dynamical entropy is a reliable estimate for the initial entropy density $s(0)$ of the heavy-ion collision.  It has been obtained by using weak coupling methods instead of the usual strong coupling holographic methods. Both the rate of entropy production and the thermalization proper-time $\tau_{th}$ are controlled by the size of the entropy density $s(0)$. In this context, the thermalization time depends on the strength of the QCD dynamical entropy of the initial dense system. Therefore, the future next step is to investigate the prediction for $s(0)$ using the  more realistic calculations of dynamical entropy  presented here (including the theoretical uncertainties).

\section{Summary}
\label{sec4}

We have studied QCD dynamical entropy for high-energy proton-nucleus collisions, as theoretically derived from the gluon dipole transverse momentum distribution (TMD) initially proposed in Ref. \cite{Peschanski:2013}. To adapt the proton transverse momentum probability distributions, we employed two strategies. The first strategy involves scaling the proton saturation scale to the nuclear case, denoted as $Q_{s}^2\rightarrow Q_{s,A}^2$, using geometric scaling properties. In these cases, we utilized analytical models for the nonintegrated gluon distribution (UGD). Specifically, we considered the MPM phenomenological model, which accurately describes charged particle spectra measured at the LHC, along with the Gaussian model (CGC Gaussian) and the model based on the Levin-Tuchin law at low-$k$ (CGC LV model). All UGDs exhibit geometric scaling properties, with their probability distribution peaks located around $k\sim Q_s$. The corresponding dynamical entropy, $\Sigma^{Y_0\rightarrow Y}(\Delta Y)$, is computed without $A$-dependence due to the standard normalization procedure in the dynamical entropy formalism and geometric scaling properties. Additionally, we computed the total entropy density, $dS_D/dy$, which shows a strong dependence on $\Delta Y$, especially noticeable for the CGC Gaussian model.

The second approach used to compute the dynamical entropy for proton-nucleus collisions involved the Glauber-Gribov framework, where we adapted the proton dipole cross-section $\sigma_{\mathrm{dip}}(r,Y)$ to the nuclear dipole cross-section $\sigma_{\mathrm{dA}}(r,Y,b)$. In this framework, we also observed $A$-independence in the nuclear dynamical entropy. However, both the nuclear transverse probability density and the total entropy generated in the CGC medium were larger than  the results obtained from the first computational strategy.

In summary, the study of nuclear QCD dynamical entropy in hadron scattering processes focused on realistic models for the dipole transverse momentum-dependent function. The analysis helped to elucidate key aspects of the rapidity dependence of the total nuclear dynamical entropy density in nuclear heavy ion collisions.  This entropy notion is related to a relative entropy and can be useful in future measurements based on quantum information theory.  The dynamical entropy can also be used to estimate the initial entropy density $s(0)$ of the heavy-ion collision based on weak coupling methods. Therefore, the thermalization time depends on the strength of the dynamical entropy of  the initial saturated state.

\begin{acknowledgments}
 MVTM acknowledges funding from the Brazilian agency Conselho Nacional de Desenvolvimento Cient\'ifico e Tecnol\'ogico (CNPq) with grant CNPq/303075/2022-8.
\end{acknowledgments}


\begin{thebibliography}{99}

\bibitem{Munier:2009pc}
S. Munier,
Phys. Rept. \textbf{473}, 1-49 (2009).
doi:10.1016/j.physrep.2009.02.001.
[\href{https://arxiv.org/abs/0901.2823}{arXiv:0901.2823 [hep-ph]}]

\bibitem{Iancu:2004es}
E. Iancu, A. H. Mueller, and S. Munier,
Phys. Lett. B \textbf{606}, 342-350 (2005).
doi:10.1016/j.physletb.2004.12.009.
[\href{https://arxiv.org/abs/hep-ph/0410018}{arXiv:hep-ph/0410018}]

\bibitem{Le:2022exz}
A. D. Le,
[\href{https://arxiv.org/abs/2203.00346}{arXiv:2203.00346 [hep-ph]}]

\bibitem{Kharzeev:2017qzs}
D. E. Kharzeev and E. M. Levin,
Phys. Rev. D \textbf{95}, 114008 (2017).

\bibitem{Kharzeev:2021nzh}
D.~E.~Kharzeev,
%``Quantum information approach to high energy interactions,''
Phil. Trans. A. Math. Phys. Eng. Sci. \textbf{380}, no.2216, 20210063 (2021)
doi:10.1098/rsta.2021.0063
[arXiv:2108.08792 [hep-ph]].


\bibitem{Hentschinski:2024gaa}
M.~Hentschinski, D.~E.~Kharzeev, K.~Kutak and Z.~Tu,
%``QCD evolution of entanglement entropy,''
Rept. Prog. Phys. \textbf{87}, no.12, 120501 (2024)
doi:10.1088/1361-6633/ad910b
[arXiv:2408.01259 [hep-ph]].

\bibitem{Hentschinski:2023izh}
M.~Hentschinski, D.~E.~Kharzeev, K.~Kutak and Z.~Tu,
%``Probing the Onset of Maximal Entanglement inside the Proton in Diffractive Deep Inelastic Scattering,''
Phys. Rev. Lett. \textbf{131}, no.24, 241901 (2023)
doi:10.1103/PhysRevLett.131.241901
[arXiv:2305.03069 [hep-ph]].

\bibitem{Hentschinski:2022rsa}
M.~Hentschinski, K.~Kutak and R.~Straka,
%``Maximally entangled proton and charged hadron multiplicity in Deep Inelastic Scattering,''
Eur. Phys. J. C \textbf{82}, no.12, 1147 (2022)
doi:10.1140/epjc/s10052-022-11122-1
[arXiv:2207.09430 [hep-ph]].


\bibitem{Zhang:2021hra}
K.~Zhang, K.~Hao, D.~Kharzeev and V.~Korepin,
%``Entanglement entropy production in deep inelastic scattering,''
Phys. Rev. D \textbf{105}, no.1, 014002 (2022)
doi:10.1103/PhysRevD.105.014002
[arXiv:2110.04881 [quant-ph]].


\bibitem{Kharzeev:2021yyf}
D.~E.~Kharzeev and E.~Levin,
%``Deep inelastic scattering as a probe of entanglement: Confronting experimental data,''
Phys. Rev. D \textbf{104}, no.3, L031503 (2021)
doi:10.1103/PhysRevD.104.L031503
[arXiv:2102.09773 [hep-ph]].

\bibitem{H1:2020zpd}
V.~Andreev \textit{et al.} [H1],
%``Measurement of charged particle multiplicity distributions in DIS at HERA and its implication to entanglement entropy of partons,''
Eur. Phys. J. C \textbf{81}, no.3, 212 (2021)
doi:10.1140/epjc/s10052-021-08896-1
[arXiv:2011.01812 [hep-ex]].


\bibitem{Ramos:2020kaj}
G. S. Ramos and M. V. T. Machado,
Phys. Rev. D \textbf{101}, 074040 (2020).
doi:10.1103/PhysRevD.101.074040.
[\href{https://arxiv.org/abs/2003.05008}{arXiv:2003.05008 [hep-ph]}]



\bibitem{Tu:2019ouv}
Z.~Tu, D.~E.~Kharzeev and T.~Ullrich,
%``Einstein-Podolsky-Rosen Paradox and Quantum Entanglement at Subnucleonic Scales,''
Phys. Rev. Lett. \textbf{124}, no.6, 062001 (2020)
doi:10.1103/PhysRevLett.124.062001
[arXiv:1904.11974 [hep-ph]].

\bibitem{Gotsman:2020bjc}
E.~Gotsman and E.~Levin,
%``High energy QCD: multiplicity distribution and entanglement entropy,''
Phys. Rev. D \textbf{102}, no.7, 074008 (2020)
doi:10.1103/PhysRevD.102.074008
[arXiv:2006.11793 [hep-ph]].

\bibitem{Germano:2021brq}
G.~R.~Germano and F.~S.~Navarra,
%``Energy dependence of the multiplicity moments at the LHC,''
Phys. Rev. D \textbf{105}, no.1, 014005 (2022)
doi:10.1103/PhysRevD.105.014005
[arXiv:2110.12028 [hep-ph]].



\bibitem{Ramos:2024}
G. S. Ramos and M. V. T. Machado,
Astronomische Nachrichten \textbf{345}, e230173 (2024).
doi:10.1002/asna.20230173.

\bibitem{Kutak:2022}
M. Hentschinski, K. Kutak, and R. Straka,
Eur. Phys. J. C \textbf{82}, 11122 (2022).
doi:10.1140/epjc/s10052-022-11122-1.
[\href{https://arxiv.org/abs/2207.09430v2}{arXiv:2207.09430 [hep-ph]}]



\bibitem{Maldacena:2000}
O. Aharony, S. S. Gubser, J. M. Maldacena,
Phys. Rept. \textbf{323}, 183 (2000).
doi:10.1016/S0370-1573(99)00083-6.
[\href{https://arxiv.org/abs/hep-th/9905111}{arXiv:hep-th/9905111}]

\bibitem{Peschanski:2013}
R. Peschanski,
Phys. Rev. D \textbf{87}, 034042 (2013).
doi:10.1103/PhysRevD.87.034042.
[\href{https://arxiv.org/abs/1211.6911}{arXiv:1211.6911 [hep-ph]}]

\bibitem{10.2996/kmj/1138844604}
H. Umegaki, Math. Sem. Rep. 14 (2) 59 - 85, 1962. https://doi.org/10.2996/kmj/1138844604.

\bibitem{Floerchinger:2020ogh}
S.~Floerchinger and T.~Haas,
%``Thermodynamics from relative entropy,''
Phys. Rev. E \textbf{102}, no.5, 052117 (2020)
doi:10.1103/PhysRevE.102.052117

\bibitem{Vedral:2002zz}
V.~Vedral,
%``The role of relative entropy in quantum information theory,''
Rev. Mod. Phys. \textbf{74}, 197-234 (2002)
doi:10.1103/RevModPhys.74.197
[arXiv:quant-ph/0102094 [quant-ph]].


\bibitem{Nielsen:2012yss}
M.~A.~Nielsen and I.~L.~Chuang,
%``Quantum Computation and Quantum Information,''
Cambridge University Press, 2012,
ISBN 978-0-521-63503-5
doi:10.1017/cbo9780511976667


\bibitem{Berges:2017hne}
J.~Berges, S.~Floerchinger and R.~Venugopalan,
%``Dynamics of entanglement in expanding quantum fields,''
JHEP \textbf{04}, 145 (2018)
doi:10.1007/JHEP04(2018)145
[arXiv:1712.09362 [hep-th]].

\bibitem{Ramos2022}
G. S. Ramos and M. V. T. Machado,
Phys. Rev. D \textbf{105}, 094009 (2022).
doi:10.1103/PhysRevD.105.094009.
[\href{https://arxiv.org/abs/2203.10986v1}{arXiv:2203.10986v1 [hep-ph]}]




\bibitem{Jarzynski:1997}
C. Jarzynski,
Phys. Rev. Lett. \textbf{78}, 2690 (1997).
doi:10.1103/PhysRevLett.78.2690.
[\href{https://arxiv.org/abs/cond-mat/9610209}{arXiv:cond-mat/9610209 [cond-mat.stat-mech]}]

\bibitem{Crooks:1999}
G. E. Crooks,
Phys. Rev. E \textbf{60}, 2721 (1999).
doi:10.1103/PhysRevE.60.2721.
[\href{https://arxiv.org/abs/cond-mat/9901352}{arXiv:cond-mat/9901352v4 [cond-mat.stat-mech]}]

\bibitem{Hatano:2001}
T. Hatano and S. Sasa,
Phys. Rev. Lett. \textbf{86}, 3463 (2001).
doi:10.1103/PhysRevLett.86.3463.
[\href{https://arxiv.org/abs/cond-mat/0010405}{arXiv:cond-mat/0010405v1 [cond-mat.stat-mech]}]

\bibitem{Mounier:2012}
A. Mounier and A. Naert,
Europhys. Lett. \textbf{100}, 30002 (2012).
doi:10.1209/0295-5075/100/30002.
[\href{https://arxiv.org/abs/1208.4039}{arXiv:1208.4039v2 [cond-mat.stat-mech]}]

\bibitem{GolecBiernat:1998}
K. J. Golec-Biernat and M. Wusthoff,
Phys. Rev. D \textbf{59}, 014017 (1998).
doi:10.1103/PhysRevD.59.014017.
[\href{https://arxiv.org/abs/hep-ph/9807513}{arXiv:hep-ph/9807513}]

\bibitem{Kutak:2011}
K. Kutak,
Phys. Lett. B \textbf{705}, 217 (2011).
doi:10.1016/j.physletb.2011.09.113.
[\href{https://arxiv.org/abs/1103.3654v4}{arXiv:1103.3654v4 [hep-ph]}]

\bibitem{Moriggi:2020zbv}
L. S. Moriggi, G. M. Peccini, and M. V. T. Machado,
Phys. Rev. D \textbf{102}, 034016 (2020).
doi:10.1103/PhysRevD.102.034016.
[\href{https://arxiv.org/abs/2005.07760}{arXiv:2005.07760 [hep-ph]}]

\bibitem{Moriggi:2020qla}
L. S. Moriggi, G. M. Peccini, and M. V. T. Machado,
Phys. Rev. D \textbf{103}, 034025 (2021).
doi:10.1103/PhysRevD.103.034025.
[\href{https://arxiv.org/abs/2012.05388}{arXiv:2012.05388 [hep-ph]}]

\bibitem{Abir:2017mks}
R. Abir and M. Siddiqah,
Phys. Rev. D \textbf{95}, 074035 (2017).
doi:10.1103/PhysRevD.95.074035.
[\href{https://arxiv.org/abs/1702.03640}{arXiv:1702.03640 [hep-ph]}]

\bibitem{Siddiqah:2018qey}
M. Siddiqah, N. Vasim, K. Banu, R. Abir, and T. Bhattacharyya,
Phys. Rev. D \textbf{97}, 054009 (2018).
doi:10.1103/PhysRevD.97.054009.
[\href{https://arxiv.org/abs/1801.01637}{arXiv:1801.01637 [hep-ph]}]

\bibitem{Kutak:2011rb}
K.~Kutak,
%``Gluon saturation and entropy production in proton\textendash{}proton collisions,''
Phys. Lett. B \textbf{705}, 217-221 (2011)
doi:10.1016/j.physletb.2011.09.113
[arXiv:1103.3654 [hep-ph]].

\bibitem{Armesto:2004}
N. Armesto, C. A. Salgado, and U. A. Wiedemann,
[\href{https://arxiv.org/abs/hep-ph/0407018}{arXiv:hep-ph/0407018}]

\bibitem{Glauber:1959}
R. J. Glauber,
Lectures in Theoretical Physics (1959).

\bibitem{Golec-Biernat:2017}
K. J. Golec-Biernat and S. Sapeta,
JHEP \textbf{03}, 102 (2018).
doi:10.1007/JHEP03(2018)102.
[\href{https://arxiv.org/abs/1711.11360}{arXiv:1711.11360 [hep-ph]}]

\bibitem{Armesto:2002}
N. Armesto,
Eur. Phys. J. C \textbf{26}, 35-43 (2002).
doi:10.1007/s10052-002-1021-z.
[\href{https://arxiv.org/abs/hep-ph/0206017v2}{arXiv:hep-ph/0206017v2}]

\bibitem{Kovner:2015hga} A. Kovner and M. Lublinsky, Phys. Rev. D. {\bf 92}, no. 3, 034016 (2015).

\bibitem{Duan:2020jkz} H. Duan, C. Akkaya, A. Kovner and V.V. Skokov, Phys. Rev. D {\bf 101}, no. 3, 036017 (2020).

\bibitem{Hagiwara:2017uaz} Y. Hagiwara, Y. Hatta, B.W. Xiao and F. Yuan, Phys. Rev. D {\bf 97}, no. 9, 094029 (2018).

\bibitem{Hatta:2016dxp}  Y.~Hatta, B.~W.~Xiao and F.~Yuan,  Phys.  Rev.  Lett.  {\bf 116}, no. 20, 202301 (2016).  

\bibitem{Kovchegov:1998bi}  Y.V. Kovchegov and A.H. Mueller,   Nucl. Phys. B {\bf 529}, 451 (1998).
  
\bibitem{Dominguez:2011wm}  F.~Dominguez, C.~Marquet, B.~W.~Xiao and F.~Yuan,  Phys. Rev. D {\bf 83}, 105005 (2011).
  
\bibitem{Rezaeian:2013tka}  A.~H.~Rezaeian and I.~Schmidt,  Phys.  Rev. D {\bf 88}, 074016 (2013).

\bibitem{Schrofl:2023hnz}
M.~Schr{\"o}fl and S.~Floerchinger,
%``Relative Entropy and Mutual Information in Gaussian Statistical Field Theory,'' 
Ann. Henri Poincaré (2024),
doi:10.1007/s00023-024-01522-2
[arXiv:2307.15548 [cond-mat.stat-mech]].

\bibitem{Nielsen2023} F. Nielsen,  Entropy \textbf{25}, 4, 654 (2023). doi:10.3390/e25040654

\bibitem{Bose:2020shm}
A.~Bose, P.~Haldar, A.~Sinha, P.~Sinha and S.~S.~Tiwari,
%``Relative entropy in scattering and the S-matrix bootstrap,''
SciPost Phys. \textbf{9}, 081 (2020)
doi:10.21468/SciPostPhys.9.5.081
[arXiv:2006.12213 [hep-th]].

\bibitem{Dowling:2020nxc}
N.~Dowling, S.~Floerchinger and T.~Haas,
%``Second law of thermodynamics for relativistic fluids formulated with relative entropy,''
Phys. Rev. D \textbf{102}, no.10, 105002 (2020),
doi:10.1103/PhysRevD.102.105002
[arXiv:2008.02706 [quant-ph]].




\end{thebibliography}
\end{document}